\documentclass{elsart}
\usepackage{psfig,natbib}
\begin{document}

\def\approxlt{\mathrel{\hbox{\rlap{\lower.55ex \hbox {$\sim$}}
        \kern-.3em \raise.4ex \hbox{$<$}}}}
\def\approxgt{\mathrel{\hbox{\rlap{\lower.55ex \hbox {$\sim$}}
        \kern-.3em \raise.4ex \hbox{$>$}}}}
\def\ros{{\sl ROSAT }}
\def\asca{{\sl ASCA }}
\def\G{$\Gamma_{\rm x}$ }

\runauthor{Stefanie Komossa}
\begin{frontmatter}
\title{Warm absorbers in Narrow-Line Seyfert\,1 galaxies}
\author[s]{Stefanie Komossa}
\address[s]{Max-Planck-Institut f\"ur extraterrestrische Physik, 
Giessenbachstr., D-85748 Garching, Germany;
email: skomossa@xray.mpe.mpg.de}
\begin{abstract}
We present a study of the X-ray properties of several
NLS1 galaxies, focusing on their warm absorbers.
In the first part, we discuss properties of 
dusty and dust-free warm absorbers in NLS1s, and study
their potential contribution to high-ionization optical iron lines.
In the second part,  
we summarize our work on the exceptional spectral
variability of the NLS1 galaxy RX J0134.3$-$4258 (from $\Gamma \simeq -4.4$
in the ROSAT all-sky survey observation, 
to $\simeq -2.2$ in our subsequent pointed observation). 

\end{abstract}
\end{frontmatter}

\section{Dusty warm absorbers in NLS1 galaxies}
Warm absorbers (WAs) are an important new probe of the
physical conditions
within the central regions of active galaxies.
They have been observed in $\sim$50\% of the well-studied Seyfert galaxies
and have also been detected in quite a number of
Narrow-line Seyfert\,1 galaxies
(see Komossa 2000 for a review, and references therein).
The photoionization calculations presented below were carried out with Ferland's 
(1993) code {\em Cloudy}.

{\em Dusty} WAs were suggested to be present in several NLS1 galaxies. 
Model calculations of dusty WAs were first applied to NGC\,4051 (Komossa \& Fink 
1997a,
KF97 hereafter). The bulk of its WA turned out to be dust-free. Other NLS1s 
were
then successfully fit with this model. Tab.\ 1 lists the results, including also
types of AGN other than NLS1s.

-------------------------------------------------------------
\begin{table*}[t]
\caption { Summary of the candidates for {\em dusty} warm absorbers, 
successfully
      modeled as such, listed in the chronological
           order that they were suggested, and results from spectral fits.
           Values of the ionization parameter $U$ given here and elsewhere in 
the text
           refer to a continuum spectrum with $\alpha_{\rm EUV}=-1.4$ (between 
Lyman-limit and 0.1 keV),
           if not noted otherwise, 
           and photon index $\Gamma_{\rm x}$ as listed. }
  \begin{tabular}{lllcll}
  \hline
   object  & type & \multicolumn{3}{l}{~~~~warm absorber fit$^{(2)}$} & 
{references$^{(3)}$} \\ 
        &  & $\Gamma_{\rm x}^{\rm intr}$ & log $U$~ & log $N_{\rm w}$ &  \\
  \hline
  \noalign{\smallskip}
    {3C\,212}  &  `red' quasar &      &       &   & [1] \\
    {IRAS\,13349+2438} & NL quasar & --2.9 & --0.4 & 21.2$^{1}$ & [2],[8] \\
    {NGC\,3227}  & Sy 1.5 &--1.9 & --0.3 & 21.8 & [3] \\
    {NGC\,3786}  & Sy 1.8 &--1.9 & --0.8 & 21.7 & [4] \\
    {MCG\,$-$6-30-15} & Sy 1 &--2.2&       & 21.7 & [5] \\
    {IRAS\,17020+4544} & NLS\,1 & --2.8 & ~0.7 & 21.6$^{1}$ & [6],[9] \\
     {4C\,+74.26} & radio quasar & --2.2 & --0.1 & 21.6 & [7]  \\
  \noalign{\smallskip}
  \hline
     \end{tabular}
  \label{tab1}

  \noindent{ \scriptsize $^{(1)}$ fixed to
     the value $N_{\rm opt}$ determined from optical reddening.
     $^{(2)}$ Comments: 3C\,212: model of dusty WA not yet fit.  
   $^{(3)}$ References: [1] Mathur 1994, [2] Brandt et al.\ 1996,
      [3] Komossa \& Fink 1997b,
      [4] Komossa \& Fink 1997c, [5] Reynolds et al.\ 1997, 
      [6] Leighly et al.\ 1997, [7] Komossa \& Meerschweinchen 2000,
      [8] Komossa \& Greiner 1999,
      [9] Komossa \& Bade 1998; for a more complete list of references 
              on the individual objects, see Komossa 2000.
     }
\end{table*}

\section{The warm absorber of NGC\,4051 and the (missing) relation between WA 
and CLR}

Dust-free warm absorbers, like the one in NGC\,4051,
might contribute significantly
to the emission in optical/UV iron coronal lines.
In the following we shall examine in detail the warm absorber
in the NLS1 galaxy NGC\,4051 with respect to the question
of whether this ionized material can account for the
optical coronal lines observed in this galaxy.
This is an update of our earlier work on this subject
(KF97).  The X-ray analysis (KF97) yielded the 
following parameters for the WA: $\log U$ = 0.4, $\log N_{\rm w}$ = 22.7,
and a density $n_{\rm{H}} \approxlt$ $3 \times 10^{7}$cm$^{-3}$,
using as input continuum the observed multi-$\lambda$ SED. 
Intensity ratios of coronal lines were taken from
the literature (e.g., 
Nagao et al.\ 2000).

We then checked, for this best-fit warm absorber model,
the strengths of the optical/UV lines originating from
the ionized material; in particular, the Fe lines
[FeVII]$\lambda$6087, [FeX]$\lambda$6374, [FeXI]$\lambda$7892,
and [FeXIV]$\lambda$5303 (see Tab.\ 2 of Komossa 2000).
We find that all of them are {\em much weaker} than observed. 
Next, we varied those parameters that do not strongly
influence the X-ray absorption structure, but
could have an effect on the predicted Fe line strengths;
namely: the metal abundances, the gas density, the EUV continuum
shape, and the IR continuum strength.
We find that in all cases, the lines [FeVII]--[FeXI] remain
much weaker than observed by several orders of
magnitude (see  Tab.\ 2 in Komossa 2000 for a detailed list of predicted
line ratios).
The reason for this is that the warm absorber is
always {\em too highly ionized}, with a totally
negligible amount of Fe$^{9+}$ and Fe$^{10+}$ ions in the gas.
Therefore, changes in collisional strengths for
the relevant Fe transitions, which are still poorly known,
are not expected to alter this result.
We conclude that for the case of NGC\,4051
{\em  the warm absorber and the coronal line region are {\sl not}
one and the same component, but are of different origins}.
This is consistent with the recent findings of Nagao
et al.\ (2000) that the [FeXI] emission of NGC\,4051 is not
confined to the nucleus, but widely
extended (out to at least $\sim$150 pc).

Recently Porquet et al.\ (1999; P99 hereafter) presented a parameter space study
of the strengths of Fe coronal lines that originate
from warm absorbers. They conclude that Fe lines
in low-density absorbers (they studied the density range
$\log n_{\rm H}$ = 8--12) are over-predicted for part of the parameter space.
Comparing our earlier results on NGC\,4051 (KF97)
with their results, we find they are consistent:
for warm absorbers dominated by OVIII absorption and high
ionization parameters, no overprediction in line emission
occurs (their Tab.\ 1).
The question remains whether the `OVII absorbers' of P99 do
indeed overpredict the Fe lines and thus are in conflict
with observations.
We propose that most of the strong OVII absorbers likely contain dust
(which was not included in the models of P99), as
suggested by the study of Reynolds (1997).
The strong depletion of Fe into dust grains then
results in weaker gas phase emission in the Fe coronal
lines; see Komossa 2000 for further discussion.  

\section{The X-ray transient NLS1 RXJ0134$-$4258}

The Narrow-line Seyfert\,1 galaxy
RXJ0134$-$4258 is one of the rare sources that showed
dramatic spectral variability. Its spectrum
changed from ultrasoft ($\Gamma_{\rm x}=-4.4$) in the {\sl ROSAT}
all-sky survey (RASS) to flat ($\Gamma_{\rm x}=-2.2$)
in our pointed
PSPC observation made two years later, while
its count rate remained nearly constant (Komossa \& Meerschweinchen 2000,
and references therein).
One possible explanation for this kind of spectral
variability is the presence of a warm absorber.

\begin{figure}[th]
\psfig{file=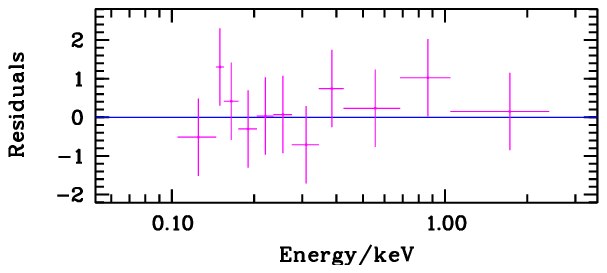,width=6.7cm,height=2.55cm,clip=}
\vspace*{-2.55cm}\hspace*{7cm} 
      \vbox{\psfig{figure=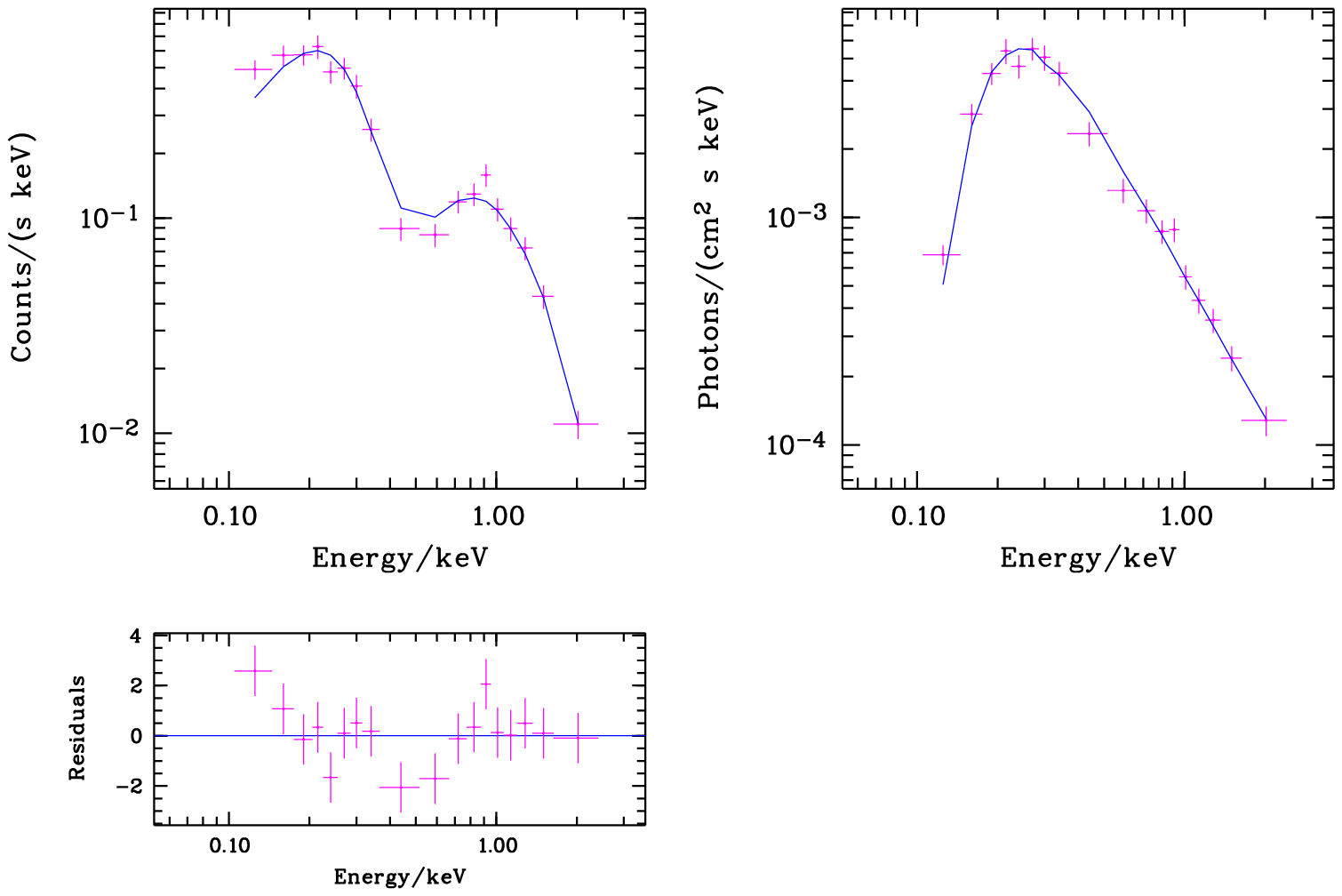,width=6.9cm,height=2.4cm,%
          bbllx=2.5cm,bblly=1.4cm,bburx=10.1cm,bbury=4.4cm,clip=}}\par
\vspace*{-0.2cm} 
\caption[poi_0134]{ {\em Left}: Residuals after fitting a {\em warm absorber} to 
the
RASS spectrum (= `{\em steep}-state' spectrum) of {RXJ0134$-$4258}.
{\em Right}: Residuals after fitting a {\em power law} to the
pointed PSPC data (= `{\em flat-state}' spectrum) of RXJ0134$-$4258.
}
\end{figure}

We find that a warmly-absorbed, intrinsically flat power law fits 
the RASS observation well, with log $N_{\rm w} \simeq 10^{23}$ cm$^{-2}$
($\chi^2_{\rm red}=0.6$). 
To account then for the much flatter spectrum during the later
PSPC observation requires a change in ionization state of the warm
absorber. Since the intrinsic luminosity of the source is not
significantly different between the two observations, it is then
required that the ionization state of the warm absorber reflects
the (unobserved) history of the variability in the intrinsic luminosity
(see KM2000 for details).
After allowing for non-equilibrium effects in the absorber and/or a range
in densities, such a warm absorber is consistent with the long-
and short-timescale variability behavior of RXJ0134$-$4258.
We did not favor this explanation, because
it introduces a new level of complexity (more free parameters) as
compared to the simpler case of an absorber in equilibrium.
Alternatively, a cloud of ionized material may have passed through 
our line of sight during the RASS observation,
and may have (nearly) disappeared during the later PSPC observation.
Finally, it is interesting to note that 
the presence of high-ionization UV absorption
lines in this object was reported by Goodrich
at this meeting.  
Indeed, recent studies suggest a nearly 
one-to-one match of the presence of UV and X-ray warm absorbers.
For a more detailed discussion of RXJ0134$-$4258, including alternative model
descriptions, we refer to Komossa \& Meerschweinchen 2000 (and references 
therein). \\
\noindent {\sl Acknowledgements:}
\noindent We thank Gary Ferland for providing {\em Cloudy}.
Preprints of this and related papers can be retrieved from \\
http://www.xray.mpe.mpg.de/$\sim$skomossa/

\end{document}